\documentclass{ws-ijmpb}

\begin{document}  
\markboth{K. Shizuya}
{Long-wavelength gauge symmetry and translations \dots}

%
\catchline{}{}{}{}{}
%

\title{Long-wavelength gauge symmetry and translations in a magnetic field\\
 for Dirac electrons in graphene}
\author{K. Shizuya}
\address{Yukawa Institute for Theoretical Physics\\
Kyoto University,~Kyoto 606-8502,~Japan\\
shizuya@yukawa.kyoto-u.ac.jp}

\maketitle

\begin{history}
\received{11 March 2019}
\accepted{2 May 2019}
Published{\  11 July 2019}
\end{history}

\begin{abstract} 
In two-dimensional (2D) electron systems in a magnetic field, the Coulomb interaction 
among charge carriers, under Landau quantization, essentially governs 
a variety of many-body phenomena while there are also phenomena, 
such as the (integer) quantum Hall effect, that appear unaffected by the interaction.
It is pointed out  that the response of 2D electrons to spatially-uniform potentials and fields 
enjoys a  long-wavelength  gauge symmetry, associated with cyclotron motion of electrons, 
that leaves the Coulomb interaction invariant 
and that thus naturally explains why cyclotron resonance (as implied by Kohn's theorem) 
and the quantized Hall conductance appear insensitive to the interaction. 
It is discussed, in the light of this new long-wavelength gauge symmetry, 
how Dirac electrons in graphene and conventional 2D electrons differ 
in cyclotron-resonance characteristics and the quantum Hall effect.

\end{abstract} 



\maketitle

\section{Introduction}

Two-dimensional (2D) electron systems  such as GaAs heterostructures\cite{PG} 
and graphene$^{2-4}$ attract great attention 
in both applications and fundamental physics 
for their novel transport characteristics.
The Coulomb interaction among charge carriers drives a variety of many-body phenomena 
and its role becomes more important in lower dimensions. 
In a magnetic field, in particular, the kinetic energy of 2D electrons is quantized 
to form a tower of Landau levels and, under this large kinetic degeneracy, 
the Coulomb interaction essentially governs the physics of many-body correlations, 
such as the fractional quantum Hall effect$^{5,6}$
and collective excitations$^{7-12}$
arising from the interplay of interaction and internal degrees of freedom (spin, valley, layer, etc).

On the other hand,  
the Coulomb interaction tends to scarcely affect long-wavelength electronic response 
such as the quantized Hall conductance 
in the quantum Hall effect (QHE)$^{13-16}$
and cyclotron resonance (CR).
In particular, Kohn's theorem\cite{Kohn} regarding the latter tells us that 
(i) CR takes place only between the adjacent Landau levels 
and (ii) the resonance energy $\omega_{c} = eB/m^{*}$ is unaffected 
by the Coulomb interaction (in the absence of disorder).

In this paper we wish to explore the principle that underlies 
such interaction-insensitive characteristics of 2D electrons in a magnetic field 
and examine its consequences.
A clue comes from a gauge symmetry encountered 
in an early study\cite{KS_gauge} of the long-wavelength response (and hence the QHE) 
of conventional 2D electrons (with quadratic dispersion) in a magnetic field.
This gauge symmetry is associated with cyclotron motion of electrons
and governs how the spatially-averaged currents 
(or total currents like $J_{x} =\int dy\, j_{x}$)
respond to spatially-uniform time-varying electric fields.
A fresh look into this gauge symmetry reveals that 
it leaves the Coulomb interaction invariant, 
that it leads to the same consequence as Kohn's theorem,\cite{Kohn} 
and that it also emerges in a purely static setting.
It is noted that translations in a magnetic field are realized in two ways, 
those in center coordinates ${\bf r}$, known as magnetic translations,$^{19,20}$ 
and those in cyclotron coordinates ${\bf X}$;
they play distinct roles in electronic transport but are related 
via electromagnetic gauge transformations.
We further extend such a long-wavelength gauge symmetry 
to Dirac electrons in graphene.
Adapting it to Dirac spinors reveals some critical differences
in response between Dirac and conventional 2D electrons.
Kohn's theorem, in particular, does not apply to electrons in graphene 
and cyclotron resonance undergoes renormalization 
while quantization of the Hall conductance remains exact 
in the presence of disorder and interaction.

In Sec.~2, we show, for conventional 2D electrons in a magnetic field, 
how one encounters, via the study of electromagnetic response, 
a long-wavelength gauge symmetry. 
A close look  is made into some characteristics of CR 
in comparison with  Kohn's theorem.
In Sec.~3, we examine two distinct types of translations in a magnetic field,
and clarify their roles in connection with disorder, localization and the resulting integer QHE.
In Sec.~4, we develop, for Dirac electrons in graphene, 
an analogous study of electromagnetic response and 
formulate a long-wavelength gauge symmetry; there we see clearly 
how conventional and Dirac fermions differ in their transport and response. 
In Sec.~5, we examine the conservation laws associated with the two types of translations
(in ${\bf X}$ and ${\bf r}$) 
and note that they neatly summarize the basic features of electronic transport in a magnetic field.
In Sec.~6, we calculate optical response of electrons in graphene and see 
how the $\lq\lq$relativistic" nature of Dirac electrons is reflected in the many-body corrections.


\section{Electrons in a Magnetic Field} 

Consider conventional 2D electrons in a magnetic field $B_{z}=B>0$, 
with the vector potential $(A_{x}, A_{y}) = (-By,0)$.
The one-body Hamiltonian 
\begin{eqnarray} 
H &=& \int dx dy \, \Psi^{\dag} {\cal H}\Psi,  \nonumber\\
{\cal H} &=&  {1\over{2m^{*}}} \{(p_{x}- eBy)^2 
+ p_{y}^2 \} = {1\over{2}}\,  \omega_{c}\, (Y^2 + P^2),
\label{H_two-d_e}
\end{eqnarray}
is essentially a harmonic-oscillator system 
with the normalized coordinate $Y=(y-y_{0})/\ell$ and momentum 
$P= \ell\, p_{y}$ with $[Y,P]= i$, where $\ell \equiv 1/\sqrt{eB}$
is the magnetic length and $y_{0}\equiv \ell^2 p_{x}$.
The electron spectrum forms Landau levels of energy 
$\epsilon_{n}= \omega_{c} (n + {1\over{2}})$ with $\omega_{c}= eB/m^{*}$,
and the eigenmodes 
$\langle x,y|n,y_{0}\rangle = \langle x|y_{0}\rangle\,  \langle y-y_{0}|n\rangle$,
labeled by $n\in (0,1,2,\cdots)$ and  $y_{0}$,
consist of plane waves $\langle x|y_{0}\rangle = e^{ix\,y_{0}/\ell^2}/\sqrt{2\pi \ell^2}$
and the harmonic-oscillator wave functions $\langle y |n\rangle$.   
In the $|n,y_{0}\rangle \equiv |{\cal N}\rangle$ basis 
the coordinate ${\bf x}=(x,y)$ is written as\cite{KS_gauge} 
\begin{eqnarray}
\langle {\cal N}| y |{\cal N}' \rangle 
&=& \{ y_{0}\, \delta^{nn'} + \ell\, Y^{nn'}\}\, \delta (y_{0}-y'_{0}),
\nonumber\\
\langle {\cal N}|x|{\cal N}'\rangle &=& 
\{   \delta^{nn'}  i\ell^2 \partial/\partial y_{0} + \ell\, P^{nn'} \}\, \delta (y_{0}-y'_{0}),
\end{eqnarray}
where $(Y, P)$ now stand for numerical matrices in level (or orbital) indices
of the familiar harmonic-oscillator form.

An electron thus undergoes relative (cyclotron) motion 
with matrix coordinate  ${\bf X} = \{X_{i}\}= \ell\, (P,Y)$ 
[with $i \in (1,2)$ or $(x,y)$] and 
center-of-mass motion 
with continuous coordinate ${\bf r} \equiv (r_{x}, r_{y}) = (i\ell^2 \partial_{y_{0}}, y_{0})$.
In what follows we make extensive use of the $|n,y_{0}\rangle$ basis, 
and denote the coordinate ${\bf x}$
as $\hat{\bf x} = {\bf r} + {\bf X}$,
with uncertainty $[X_{x}, X_{y}] =-i\ell^{2}$, $[r_{x}, r_{y}] =i\ell^{2}$ 
and $[X_{i},r_{j}]=0$.

To study the electromagnetic response of the system 
let us here introduce external potentials $v_{\mu} = (v_{x},v_{y}, v_{0})$. 
They are taken to be spatially-uniform 
but slowly varying in time. 
Actually it suffices to employ such long-wavelength potentials $v_{\mu}(t)$
to study the basic transport property of the system:
They serve to detect, e.g., the total current $\int dy\,j_{x}$ (or the $x$-averaged one 
$(1/L_{x}) \int dxdy\, j_{x}$ with $L_{x}= \int dx$)
driven by an applied electric field 
$(-\partial_{t} v_{x}, -\partial_{t} v_{y})$.

Passing to the $|n,y_{0}\rangle$ basis via the expansion 
$\Psi ({\bf x}) = \sum_{n, y_{0}} \langle {\bf x}|n,y_{0}\rangle\, \psi^{n}(y_{0})$
yields the Hamiltonian
\begin{eqnarray}
&&H = \int dy_{0}\, \sum_{m,n}\psi^{m \dag}(y_{0})\, {\cal H}[v,v_{0}]^{mn}\, \psi^{n}(y_{0}),
\\
&&{\cal H}[v,v_{0}] = \omega_{c}
\left\{ (Z^{\dag} -i v^{\dag}) (Z+ i v) + 1/2  \right\} - e v_{0} ,
\end{eqnarray}
where 
$v \equiv e\ell\, (v_{y}+ iv_{x})/\sqrt{2}$ and 
$v^{\dag} \equiv e\ell\, (v_{y}- iv_{x})/\sqrt{2}$;
$Z \equiv (Y+ iP)/\sqrt{2}$ and  $Z^{\dag} \equiv (Y- iP)/\sqrt{2}$
with $[Z,Z^{\dag}]=1$; 
$Z^{mn} \equiv \langle m| Z| n\rangle = \sqrt{n}\, \delta^{m,n-1}$
in the standard notation.
Obviously $v$ and $v_{0}$ are diagonal in level indices, 
with the unit matrix  $1 \sim \delta^{mn}$ suppressed.
For $v\not=0$, ${\cal H}[v,v_{0}]$ is no longer diagonal.
In what follows, for notational clarity, 
we adopt matrix notation and frequently suppress summation over level indices.

In the $|n,y_{0}\rangle$ basis the charge density 
$\rho_{-{\bf p}} =\int d^{2}{\bf x}\,  e^{i {\bf p\cdot x}}\,\rho$ 
with $\rho = \Psi^{\dag} \Psi$ 
is written as
\begin{eqnarray}
\rho_{-{\bf p}} &=&
 \int dy_{0}\, \psi^{\dag} e^{i{\bf p\cdot \hat{x}}}\, \psi 
=\sum_{m, n} U^{m n}_{\bf p}\, R^{m n}_{-{\bf p}}, 
\\
R^{mn}_{ -{\bf p}}&\equiv& \int dy_{0}\,
\psi^{m \dag}(y_{0})\, e^{i{\bf p\cdot r}}\,
\psi^{n}(y_{0}), 
\label{Rmnp}
\\
U_{\bf p} &\equiv&  e^{i{\bf p\cdot X}} =e^{ i(p Z^{\dag} + p^{\dag}Z)}
\equiv U_{p},
\label{U_p}
\end{eqnarray}
where $U^{mn}_{\bf p} \equiv \langle m|U_{\bf p} |n\rangle$; 
$p\equiv \ell\, (p_{y} + ip_{x})/\sqrt{2}$
and $p^{\dag}\equiv \ell (p_{y} - ip_{x})/\sqrt{2}$.
Remember that we denote 
$U_{\bf p}$ as $U_{p}$ by replacing 
the suffix ${\bf p} =(p_{x},p_{y})$ with the dimensionless complex suffix $p$.
Here the charge operators $R^{mn}_{ -{\bf p}}$ obey the $W_{\infty}$ algebra\cite{GMP} 
or the composition law $e^{i {\bf p\cdot r}} e^{i {\bf k\cdot r}} 
= e^{-i{1\over{2}} \ell^{2}  {\bf p \times k}}\, e^{i {\bf (p+k)\cdot r} }$
(with ${\bf k\! \times\! p} \equiv k_{x}p_{y}-k_{y}p_{x}$),
 that reflects the uncertainty $[r_{x}, r_{y}] =i\ell^{2}$ of ${\bf r}$.
The form-factor matrices $U_{\bf p}$ also obey the $W_{\infty}$ algebra,
\begin{eqnarray}
&&U_{\bf p}\, U_{\bf q} =U_{p+q}\,  e^{ {1\over{2}}(p\, q^{\dag} -p^{\dag}q)}
= U_{\bf p+q}\,  e^{i {1\over{2}} \ell^2 {\bf p} \times {\bf q}},
\\
&&U_{\bf q}\,  U_{\bf p} U_{\bf-q} = e^{\, q\, p^{\dag} - q^{\dag}p}\,  U_{\bf p}
= e^{- i \ell^2 {\bf p} \times {\bf q} }\, U_{\bf p}.
\label{U_complaw}
\end{eqnarray}
One can rewrite  $U_{\bf p} = e^{-{1\over{4}} \ell^2 {\bf p}^2} f_{\bf p}$
and $f_{\bf p} = e^{ i p\, Z^{\dag}} e^{ i p^{\dag} Z}$.
Then $f^{m n}_{\bf p} \equiv \langle m|f_{\bf p}|n\rangle$ are explicitly written as 
\begin{equation}
f^{m n}_{\bf p} = 
\sqrt{n!/m!}\, (ip)^{m-n}\, L^{(m-n)}_{n}(p^{\dag}p)
\label{f_mn}
\end{equation}
for $m \ge n\ge0$, and $f^{n m}_{\bf p} = (f^{m n}_{\bf -p})^{\dag}$;
$p = \ell (p_{y}+ip_{x})/\sqrt{2}$.

Finally the Coulomb interaction is denoted as
\begin{equation}
V[\rho] = {1\over{2}} \sum_{\bf p} v^{\rm C}_{\bf p} :\rho_{\bf -p}\, \rho_{\bf p}: \ ,
\end{equation}
with the potential $v^{\rm C}_{\bf p}= 2\pi \alpha/(\epsilon_{\rm b} |{\bf p}|)$,
$\alpha \equiv e^{2}/(4 \pi \epsilon_{0})$ and 
the substrate dielectric constant $\epsilon_{\rm b}$;
$\sum_{\bf p} \equiv \int d^{2}{\bf p}/(2\pi)^{2}$ and we set 
$\delta_{\bf p,0}\! \equiv (2\pi)^2 \delta^{2}({\bf p})$;
normal ordering stands for 
$: (\psi^{m \dag} \psi^{n}) (\psi^{m'\dag}\psi^{n'}): \, 
\sim \psi^{m'\dag} \psi^{m\dag} \psi^{n} \psi^{n'}$.

The full system or the Lagrangian
\begin{equation}
L = \int dy_{0}\, \psi^{\dag} 
\big\{  i\partial_{t}  - {\cal H}[v,v_{0}] \big\}\, \psi - V[\rho],
\label{full_Lag}
\end{equation}
has an interesting gauge symmetry.\cite{KS_gauge}
Consider the following unitary transformation that mixes infinitely many Landau levels,
\begin{eqnarray}
\psi_{\theta}^{m}(y_{0}) &=& (U_{\theta})^{mn}\,  \psi^{n}(y_{0}),
\nonumber\\
U_{\theta} &=& e^{i(\theta_{x} P + \theta_{y} Y )} 
= e^{ i(\theta^{\dag} Z + \theta Z^{\dag})},
\label{psi_theta}
\end{eqnarray}
where 
spatially-uniform real phases $\theta_{i} =\theta_{i}(t)$
can vary in time; 
$\theta = (\theta_{y} + i \theta_{x})/\sqrt{2}$
and $\theta^{\dag} = (\theta_{y} - i \theta_{x})/\sqrt{2}$.
This $U_{\theta}$ works to shift the relative coordinate,  
\begin{equation}
U_{\theta}{\bf X}\,  U_{\theta}^{-1} = {\bf X} - \ell\, \tilde{\theta},
\end{equation}
with $\tilde{\theta} \equiv  (\theta_{y}, -\theta_{x})$,
or  $(Z, Z^{\dag}) \rightarrow (Z- i\theta, Z^{\dag}+ i\theta^{\dag})$.
The charge thereby undergoes, in view of Eq.~(\ref{U_complaw}), 
only a phase change
\begin{equation}
\rho_{\bf -p} =e^{ -i \ell {\bf p} \cdot  \tilde{\theta}}\!
\int\!\! dy_{0}\, \psi_{\theta}^{\dag}\, U_{p} \,   e^{i{\bf p \cdot r} } \psi_{\theta}
\equiv e^{ -i \ell {\bf p}  \cdot  \tilde{\theta} }\, (\rho_{\theta})_{\bf -p},
\label{MTcharge}
\end{equation}
and the Coulomb interaction remains invariant in form,
\begin{equation}
V[\rho]= V[\rho_{\theta}]
=  {1\over{2}} \sum_{\bf p} v^{\rm C}_{\bf p} :(\rho_{\theta})_{\bf -p}\, (\rho_{\theta})_{\bf p}: .
\end{equation}
Time evolution of $U_{\theta}$ gives rise to Berry's phase,\cite{Berry}
$U_{\theta} i\partial_{t} U_{-\theta} = \dot{\theta}\, Z^{\dag}+ \dot{\theta}^{\dag} Z
- i {1\over{2}} (\theta\, \dot{\theta}^{\dag} - \theta^{\dag} \, \dot{\theta})$;
$\dot{\theta} \equiv \partial_{t} \theta$, etc.
The Lagrangian then retains the same form 
\begin{equation}
L = \int dy_{0}\, \psi_{\theta}^{\dag} 
\big\{  i\partial_{t}  - {\cal H}[v^{\theta},v_{0}^{\theta}] \big\}\, \psi_{\theta} 
- V[\rho_{\theta}],
\label{full_Lag_theta}
\end{equation}
under the transformation 
$\psi\rightarrow \psi_{\theta}$, $v\rightarrow v^{\theta}$ and 
$v_{0}\rightarrow v_{0}^{\theta}$, with
\begin{eqnarray}
v^{\theta} &=& v - \theta + (i/\omega_{c})\, \dot{\theta},
\nonumber\\
e\,  v_{0}^{\theta} &=&  e\,  v_{0}
+ i ( \dot{\theta}\, v^{\theta \dag} - \dot{\theta}^{\dag} v^{\theta} )
- i \textstyle{1\over{2}}\, 
(\theta\, \dot{\theta}^{\dag} - {\theta}^{\dag}\dot{\theta})
- (1/\omega_{c})\, \dot{\theta}^{\dag} \dot{\theta}.
\label{v_theta}
\end{eqnarray}
This invariance implies that 
the present electron system $\{ \psi \}$ in applied potentials $v_{\mu}$ 
has physically the same property 
as the transformed system  $\{ \psi_{\theta} \}$ in the potentials $v_{\mu}^{\theta}$.
We explore the physical origin of this (long-wavelength) gauge symmetry later
and here look into its  consequences.

(I) Let us first choose $\theta$ so that $v^{\theta}=0$, i.e., 
\begin{equation}
\theta = \{1- (i\partial_{t}/\omega_{c}) \}^{-1} v \equiv \theta'.
\label{choice_exact}
\end{equation}
This achieves diagonalization of the one-body part of $L$, 
\begin{equation}
L = \int dy_{0}\, \psi_{\theta'}^{\dag} 
\big\{  i\partial_{t}  - \epsilon  + e\, v_{0}^{\theta'} \big\}\, \psi_{\theta'} 
- V[\rho_{\theta'}],
\label{L_psi_theta}
\end{equation}
with $\epsilon \rightarrow \epsilon_{n}\, \delta^{mn}$, 
$\epsilon_{n} \equiv  \omega_{c}\, (n+ 1/2)$ and 
\begin{eqnarray}
e\, v_{0}^{\theta'} &=& e\, v_{0}
- i {1\over{2}}  \big\{ v\, {\omega_{c}\over{\omega_{c} 
+ i\partial_{t}}}\,   \dot{v}^{\dag}
- v^{\dag}{\omega_{c}\over{\omega_{c}- i\partial_{t}}}\,   \dot{v} 
 \big\}, \
\label{HC_response_exact}
\\
&=& e\, v_{0}
-i {\textstyle{1\over{2}} }  (v\, \dot{v}^{\dag} - v^{\dag} \dot{v} )
+ O(\partial_{t}^{2}).
\label{response-electron}
\end{eqnarray}
From the Chern-Simons term
\begin{equation}
- i \textstyle{1\over{2}} ( v\, \dot{v}^{\dag} -v^{\dag} \dot{v} )
=  {\textstyle{1\over{2}}} e^2 \ell^2 (v_{x} \dot{v}_{y} - v_{y} \dot{v}_{x}),
\end{equation}
one can read the Hall conductance $\sigma_{xy}$ 
equal to $-e^2 \ell^2$ per electron (per unit area)
or  $\sigma_{xy}=-e^2 \ell^2 \bar{\rho}
=- e^2/(2\pi \hbar) = -e^2/h$ per filled level, with level density 
$\bar{\rho} \equiv 1/(2\pi \ell^2)$.
Actually, Eq.~(\ref{HC_response_exact}) tells us more:
Varying the Lagrangian~(\ref{L_psi_theta}) with respect to $v^{\dag}$ yields
the total current operator in this $\psi_{\theta'}$ representation,
\begin{equation}
\int\! dy_{0}( j_{y}+ ij_{x})_{\theta'} 
=  ie\ell {\omega_{c}\over{\omega_{c} - i\partial_{t}}}\, 
(\dot{v}_{y}+i \dot{v}_{x}) \int\! dy_{0}\,  \psi_{\theta'}^{\dag}\psi_{\theta'}.
\label{current_resp}
\end{equation}
Here the total current is proportional to the {\it conserved} charge 
$\int dy_{0}\, \psi_{\theta'}^{\dag}\psi_{\theta'}
= \int dy_{0}\, \psi^{\dag}\psi = \int d^{2}{\bf x}\, \Psi^{\dag}\Psi$,
whose expectation value 
$\int d^{2}{\bf x}\,  \langle \Psi^{\dag}\Psi\rangle = N_{e}$, 
the total number of electrons, is unaffected by loop corrections. 
Thus the current response~(\ref{current_resp})  
is an exact one and is not corrected by the Coulomb interaction $V[\rho_{\theta'}]$.
Actually, Eq.~(\ref{HC_response_exact}) or (\ref{current_resp}) represents a CR 
of excitation energy $\omega_{c}$ (at zero-momentum transfer ${\bf p}=0$)
to the adjacent level, independent of the Coulomb interaction
and in agreement with Kohn's theorem.\cite{Kohn}
This leads to the (exact) optical conductivity 
\begin{equation}
\sigma_{xy}(\omega) = \sigma_{xy}/\{1- (\omega/\omega_{c})^2\}
\end{equation}
at finite frequency $\omega \sim i \partial_{t}$.
A direct calculation, indeed, shows that 
the resonance energy stays to be $\omega_{c}$ 
as a result of cancellation between the self-energy corrections 
and attraction energy of the created electron-hole pair.\cite{KH}

It is crucial in the above analysis that we handle the total current 
and that, in the $\psi_{\theta'}$ system~(\ref{L_psi_theta}), 
all the reference to vector potential $v$ is assembled into the $e\, v_{0}^{\theta'}$ term, 
which is coupled to the conserved charge
$\int\! dy_{0}\,  \psi_{\theta'}^{\dag}\psi_{\theta'}$.
Actually, one can equally well handle a current density $j_{i}({\bf x})$ 
coupled to a local potential $v({\bf x})$ and again remove the $O(v)$ term 
from $L$ by a suitable rotation $\psi \rightarrow \psi'$.
The current density then depends on how the electrons, 
driven by a local electric field $\propto -\dot{v}({\bf p})$, 
mutually interact via the Coulomb pontential $V[\rho]$.

(II) An alternative choice of $\theta$ is to simply set $\theta =v$, or $\theta_{i} = e \ell\, v_{i}$.
The Lagrangian then takes the form
\begin{eqnarray}
L &=& \int dy_{0}\, \psi_{v}^{\dag} \{ i\partial_{t} - {\cal H}_{v} \}\, \psi_{v} - V[\rho_{v}],
\nonumber\\
{\cal H}_{v} &=& 
\omega_{c}\, ( Z^{\dag} Z+ 1/2)
- (\dot{v} Z^{\dag}+ \dot{v}^{\dag} Z)  
- e\,  v_{0} + i \textstyle{1\over{2}} (v\, \dot{v}^{\dag} -v^{\dag}\dot{v} ), 
\label{H_v}
\end{eqnarray}
with $\psi_{v} \equiv \psi_{\theta=v}$. 
Here the Hall field $- (\dot{v} Z^{\dag}+ \dot{v}^{\dag} Z)$ 
still induces level mixing. Its effects, if calculated perturbatively,
necessarily involve two powers of $\partial_{t}$ or more 
$\sim O(\dot{v}\, \dot{v}^{\dag})$, 
and the correct value of $\sigma_{xy}$ is still read from ${\cal H}_{v}$.
For the optical response $\sigma_{xy}(\omega)$ 
one has to rotate $\psi_{v}$ slightly more, as seen from $\theta'$ in Eq.~(\ref{choice_exact}).

The long-wavelength gauge symmetry also emerges in a purely static setting, 
i.e., in studying a response to a static electric field ${\cal E}_{i}=-\partial_{i}a_{0}$;
$\dot{\cal E}=\dot{v}=0$.
Let us promote $v_{0}$ to $a_{0}({\bf x}) = v_{0} - x_{i}{\cal E}_{i}$.
Via the transformation $\psi_{\theta}= U_{\theta} \psi$, 
the one-body Hamiltonian 
$H= \int\! dy_{0}\,  \psi^{\dag} {\cal H}[v, a_{0}(\hat{\bf x})]\, \psi$ is rewritten as   
\begin{equation}
H = \int\! dy_{0}\,  \psi_{\theta}^{\dag}\, 
{\cal H}[v^{\theta}, a_{0}(\hat{\bf x} - \ell\, \tilde{\theta})]\, \psi_{\theta},
\label{psi-theta-azero}
\end{equation}
where $v^{\theta} = v- \theta$ and $ \tilde{\theta} \equiv (\theta_{y}, -\theta_{x})$.
One can rearrange ${\cal H}[v^{\theta}, \cdots]$ in the form 
${\cal H}[v^{\theta}_{\cal E}, a^{\theta}_{0}(\bf r)]$ with
\begin{eqnarray}
v^{\theta}_{\cal E}&=& v - ie\ell\, {\cal E}/\omega_{c} -\theta,
\nonumber\\
a^{\theta}_{0}({\bf r}) &=& a_{0}({\bf r} - \ell\, \tilde{\theta})
- (\omega_{c}/e) \{ |v^{\theta}|^2 -   |v^{\theta}_{\cal E}|^2  \},
\end{eqnarray}
where ${\cal E} = ({\cal E}_{y} + i {\cal E}_{y})/\sqrt{2}$.
Choosing $v^{\theta}_{\cal E} = 0$,
or $\theta = v - ie\ell\, {\cal E}/\omega_{c}$, then allows one to diagonalize
the one-body Hamiltonian in the form
\begin{equation}
 {\cal H}[0, a_{0}({\bf r} - e\ell^2\, \tilde{\bf v})] 
- e^2\ell^2 {\cal E}_{i}^{2}/(2\omega_{c}).
\label{H_st}
\end{equation}
where $\tilde{\bf v} \equiv (v_{y}, -v_{x})$.
From 
\begin{equation}
a_{0}({\bf r} - e\ell^2 \tilde{\bf v}) 
= a_{0}({\bf r}) + i \ell (v\, {\cal E}^{\dag} -v^{\dag} {\cal E} )
=  a_{0}({\bf r}) - e \ell^2 (v_{x} {\cal E}_{y} - v_{y} {\cal E}_{x} )
\end{equation}
one can again read the current driven by ${\cal E}_{i}$ and 
Hall conductance $\sigma_{xy} = -e^2/h$ per filled level.
This value of $\sigma_{xy}$ is no longer corrected by the Coulomb interaction.

\section{Translations in Center Coordinates and Localization}

In this section we explore the origin and basic role of the long-wavelength gauge symmetry. 
To this end, we consider the $\psi_{\theta}=U_{\theta}\psi$ static system 
of Eq.~(\ref{H_st})  (with $v_{0}\rightarrow 0$).
For weak field $e|{\cal E}| \ll \omega_{c}$ and to $O({\cal E})$, 
one can simply take 
\begin{equation}
{\cal H}[0,a_{0}(\hat{\bf x}-\ell\, \tilde{\theta})] 
 \approx  \epsilon  -e\,  a_{0}({\bf r}-\ell\, \tilde{\theta}) 
 \equiv {\cal H}^{\rm static}(\bar{\bf r}),
\end{equation} 
where 
$\bar{\bf r} \equiv  {\bf r} - \ell\,  \tilde{\theta} 
= (r_{x} -\ell \theta_{y}, r_{y} + \ell \theta_{x})$ 
and $\theta_{i} = e\ell\, v_{i}$.
In this section we use $\theta$ rather than the chosen value $\theta=v$ 
to emphasize its character as a transformation parameter.  
In addition, for clarity of exposition, 
we use only $\theta_{x}= e \ell v_{x}$ 
to detect the current $j_{x}$ driven by a static field ${\cal E}_{y}$;
accordingly, we denote ${\cal H}^{\rm static}(\bar{\bf r}) = \epsilon(\bar{y}_{0})$ 
with $\epsilon(y_{0}) \equiv \epsilon + e{\cal E}_{y}\, y_{0}$ 
and $\bar{y}_{0}=y_{0}+ \ell\, \theta_{x}$; 
for later generalization, however, we adopt notation with both $(v_{x}, v_{y})$.

The $U_{\theta}$ shifts potential $a_{0}(\hat{\bf x})$ in $\hat{\bf x}$ (or ${\bf X})$ 
while the electron field 
$\psi_{\theta} (y_{0}) = U_{\theta}\, \psi (y_{0})$ remains spatially unshifted.
[Note here that the $|n,y_{0}\rangle$ base, 
$\langle {\bf x}|n,y_{0}\rangle = \langle x|y_{0}\rangle\,  \langle y-y_{0}|n\rangle$,
is spatially localized around $y \sim y_{0}$ with spread $\Delta y \sim O(\ell)$ 
while it is a plane wave extended in $x$.]
Since both ${\bf X}$ and ${\bf r}$ obey the $W_{\infty}$ algebra,
it is also possible to spatially shift $a_{0}(\hat{\bf x})$ 
by translations in the center coordinate
${\bf r} = (i\ell^2 \partial_{y_{0}}, y_{0})$, 
known as magnetic translations.\cite{MT_B,MT_Zak}  
Actually, with translation
\begin{equation}
\Xi = e^{i(\theta_{x}\, r_{x} + \theta_{y}\, r_{y})/\ell}
=  e^{i{1\over{2}} \theta_{x}\theta_{y}} e^{i\theta_{x}r_{x}/\ell}\, e^{i\theta_{y} y_{0}/\ell},
\end{equation}
one can formally eliminate from ${\cal H}^{\rm static}(\bar{\bf r})$
its reference to $\theta_{i}=e\ell\, v_{i}$,
\begin{equation}
\Xi\, {\cal H}_{v}^{\rm static}(\bar{r}_{x}, \bar{r}_{y})\Xi^{-1}=  {\cal H}_{v}^{\rm static}(r_{x}, r_{y}).
\label{XiHXi}
\end{equation}
This appears to imply that the transformed field $\psi_{\xi} \equiv \Xi\, \psi_{\theta}$
carries no current driven by ${\cal E}_{y}$.
This, of course, is not the case. Let us examine this point below.

 The translations in ${\bf X}$ and in ${\bf r}$ differ in their range.
Cyclotron motion is always localized in space with $|{\bf X}| \sim O(\ell)$ bounded,
and the harmonic-oscillator eigenmodes $\phi_{n}(y-y_{0})=\langle y-y_{0}|n\rangle$ 
are normalizable (i.e., square integrable) functions and span a Hilbert space. 
The $U^{mn}_{\theta}$ are unitary rotations in this space 
and leave the energy spectra unchanged
in passing from $\psi$ to $\psi_{\theta}=U_{\theta}\psi$.
 
In contrast, center motion of orbiting electrons is not bounded since 
${\bf r} = (i\ell^2 \partial_{y_{0}}, y_{0})$ can be as large as the sample size.
Consider, e.g., a plane-wave eigenmode of ${\cal H}^{\rm static}(\bar{\bf r})$,  
$\langle y_{0}|p'_{x}\rangle = \delta (y_{0} - \ell^2 p'_{x})$ 
of (conserved) momentum $p_{x}= p'_{x}$ 
and energy 
$\epsilon(\ell^2 p'_{x} + \ell \theta_{x}) 
= \epsilon_{n} + e {\cal E}_{y}\, (\ell^2 p'_{x}+ \ell\, \theta_{x})$.
It is spatially localized about $y\sim y_{0}=\ell^2 p'_{x}$ with $|\Delta y| \sim O(\ell)$
and extended in $x$.
Such extended modes are not normalizable in $y_{0}$ or in ${\bf x}$.  
The translation $\Xi$ turns $\psi_{\theta}$ to $\psi_{\xi}=\Xi\, \psi_{\theta}$,
i.e., 
$\psi_{\xi}(y'_{0}) 
= \sum_{y_{0}} \langle y'_{0}| \Xi |y_{0}\rangle\, \psi_{\theta}(y_{0})$, 
 or explicitly, 
  \begin{equation}
 \psi_{\xi}^{n} (y_{0})=e^{i{1\over{2}} \theta_{x}\theta_{y}}\, 
 e^{i\theta_{y} (y_{0} -\ell\theta_{x})/\ell}\, \psi_{\theta}^{n} (y_{0}-\ell \theta_{x}).
 \end{equation}
Via $\Xi$, $\psi_{\theta}$ is shifted in ${\bf r}$ by $\ell\, \tilde{\theta}$.
The electron mode $\psi_{\theta} (y_{0})$  [of energy $\epsilon (\bar{y}_{0})$] 
localized around $y\sim y_{0}$ in the real space thereby turns into 
$\psi_{\xi}^{n} (y_{0}+ \ell \theta_{x})$ [of the same spectrum $\epsilon (y_{0}+ \ell \theta_{x})$] 
localized around $y \sim y_{0}+ \ell\, \theta_{x}$.
It is now clear that Eq.~(\ref{XiHXi}) does not mean the absence of 
$\theta_{i} = e\ell v_{i}$ from the spectra of $\psi_{\xi}= \Xi\, \psi_{\theta}$.
Actually, spatial shift  is precisely the way the extended modes respond 
when one turns on magnetic flux $\theta_{x}$ adiabatically, 
as noted by Laughlin\cite{Laughlin_qhe}  in his explanation for the integer QHE.
It is not a coincidence that  $U_{\theta}$ and $\Xi$ combine to form a gauge transformation 
\begin{equation}
\Lambda \equiv \Xi\,  U_{\theta}= e^{i \theta_{j} (X_{j}+r_{j})/\ell}= e^{i \theta_{j} \hat{x}_{j}/\ell},
\end{equation}
that, upon  $\psi_{\Lambda}= \Lambda\, \psi$, shifts $v \rightarrow v- \theta$ while
$a_{0}({\bf x})$ is left unchanged.

It is enlightening to see how the energy changes via a shift in $y_{0}$.
Varying $\theta_{x}$ slightly by $\delta \theta_{x}$ in Eq.~(\ref{XiHXi}) yields
$\delta {\cal H}_{v}^{\rm static}(\bar{\bf r})
= \Xi^{-1} [{\cal H}_{v}^{\rm static}({\bf r}) , \delta \Xi\, \Xi^{-1}]\, \Xi$.
The associated change of the energy
$\delta H=  \int dy_{0}\, 
\psi^{\dag}_{\theta}\, \delta {\cal H}_{v}^{\rm static}(\bar{\bf r})\, \psi_{\theta}$ 
is thereby rewritten as
\begin{equation}
\delta H= \ell\, \delta \theta_{x} \int\! dy_{0}\, 
\psi^{\dag}_{\xi}\,  [\partial_{y_{0}}{\cal H}_{v}^{\rm static}({\bf r})]\, \psi_{\xi}. 
\label{deltaH_xi}
\end{equation}
For a sample of size $L_{x} \times L_{y}$, 
a filled Landau level has  degeneracy 
$L_{x}/(2\pi \ell^2)\sim \langle \psi^{\dag}_{\xi}\psi_{\xi} \rangle$ in $y_{0}$,
and Eq.~(\ref{deltaH_xi}) tells us that the energy change
\begin{equation}
\langle \delta H \rangle = {L_{x}\over{2\pi \ell}}\, \delta \theta_{x}\! \int\! dy_{0}\, 
\partial_{y_{0}}\epsilon_{n}(y_{0})
=\delta \theta_{x}\, {L_{x}L_{y}\over{2\pi \ell}}\, e {\cal E}_{y} 
\label{deltaH_xi}
\end{equation}
is only associated with the electron modes 
that come in or go out through the sample edges ($y=0, L_{y}$).
Setting $\delta \theta_{x}\rightarrow e\ell\, \delta v_{x}$, 
one can read the current 
$\int dy\, \langle j_{x}\rangle$ driven by ${\cal E}_{y}$
and  $\sigma_{xy}$ equal to $-e^2/h$ per filled level. 
This value of $\sigma_{xy}$ is left unaffected by the Coulomb interaction $V[\rho]$, 
which is invariant under translations $U_{\theta}$ and $\Xi$.
As is clear now, this conclusion holds not only for $V[\rho]$ 
but also for general translation-invariant interactions.

The response of Hall electrons changes considerably in the presence of disorder.
Consider, as a simple example, 
a single impurity with a delta-function potential of strength $\lambda$,
\begin{equation}
V^{\rm imp}({\bf x}) = \lambda\,  \delta (x-\xi) \delta (y-\eta),
\end{equation} 
located at ${\bf x}=(\xi, \eta)$ in a sample; 
we set $-ea_{0}(\hat{\bf x}) \rightarrow -ea_{0}(\hat{\bf x}) + V^{\rm imp}(\hat{\bf x})$ 
in ${\cal H}^{\rm static}$. 
This impurity captures electrons and
there arises one localized mode [of spread $|\Delta {\bf x}| \sim O(\ell)$] 
in each Landau level $n$, with a normalizable wave function 
(in the $\psi_{\theta}$ representation) of the form,
\begin{equation}
\psi^{\rm loc}_{n} (y_{0})=e^{-i\xi'\, y_{0}/\ell^2}  \phi_{n}(y_{0}-\eta') 
\end{equation}
to $O(\lambda)$ and for weak field $e \ell |{\cal E}_{y}| \ll \lambda/(2\pi \ell^2)$;
$\phi_{n}(y) = \langle y|n\rangle$ denote the harmonic-oscillator eigenfunctions.
Here $\xi'=  \xi +\ell \theta_{y}$ and $\eta'= \eta - \ell \theta_{x}$ 
are due to the shifted ${\bf r}$ in ${\cal H}^{\rm static}(\bar{\bf r})$. 
The eigenvalue, however, is independent of the shift $\theta_{i}$,
\begin{equation}
\epsilon^{\rm loc}_{n} = \epsilon_{n} +  e{\cal E}_{y}\, \eta+ \lambda/(2 \pi \ell^{2}),
\end{equation}
which implies that such a localized mode carries no current. 
The transformation $\Xi = e^{i(\theta_{x}\, r_{x} + \theta_{y}\, r_{y})/\ell}$, 
acting on $\psi^{\rm loc}_{n} (y_{0})$,
recovers, apart from a global phase, the localized mode unshifted in $\theta_{i}$
of ${\cal H}^{\rm static}({\bf r})$,
\begin{equation}
\Xi\, \psi^{\rm loc}_{n} (y_{0}) =
e^{i \theta_{x}(\xi/\ell  + {1\over{2}}\,\theta_{y}) } e^{-i\xi\, y_{0}/\ell^2}  \phi_{n}(y_{0}- \eta),
\end{equation}
with the same eigenvalue $\epsilon^{\rm loc}_{n}$.

With more impurities there arise many localized modes in each level $n$.
They, being spatially localized, naturally have normalizable wave functions and 
span a Hilbert space within the full $(n,y_{0})$ space. 
For such normalizable modes with localized coordinates $\langle r_{i} \rangle$,
$\Xi$ acts as a well-defined unitary transformation associated with a change of bases 
[from $\{\psi^{\rm loc}_{n}\}$ to $\{\Xi\, \psi^{\rm loc}_{n} \}$ in the above example] 
and  leaves their spectra unchanged. The relation~(\ref{XiHXi}) then generally reveals
that the localized modes carry no current.
Physically this is because the localized modes, 
unlike extended ones, are insensitive to the sample boundaries 
and hence to a shift in $y_{0}$.

The Hamiltonian ${\cal H}[v, a_{0}(\hat{\bf x}) ]$ for $\psi$ turns into
${\cal H}[v^{\theta}, a_{0}(\hat{\bf x}-\ell \tilde{\theta}) ]$ 
for $\psi_{\theta}=U_{\theta}\psi$, 
and into ${\cal H}[v^{\theta}, a_{0}(\hat{\bf x})]$ 
for $\psi_{\Lambda}= \Lambda \psi$.
Translations in  ${\bf X}$, $U_{\theta}$, shift potentials $\{v, a_{0}({\bf x}) \}$ and 
induce level mixing of the electron field $\psi$ 
while $\psi_{\theta}$ remains spatially unshifted.
They thus provide a direct way to diagonalize the spectra and 
long-wavelength response of the electrons.
The gauge transformation $\Lambda= \Xi U_{\theta}$ 
can also shift away $v$.
The electron field $\psi_{\xi}= \Lambda\, \psi$ is thereby spatially shifted
and appears to carry no current.
The correct amount of current is recovered by shifting $\psi_{\xi}$ back 
to $\psi_{\theta}$ via $\Xi$. 
In this way,  translations $U_{\theta}$ and magnetic translations $\Xi$ 
are distinct in concept, though they are related 
via gauge transformations $\Lambda=\Xi\, U_{\theta}$.

Incidentally, it is worth noting here that, 
when the potential $a_{0}(\hat{\bf x})$ has a finite periodicity of ${\bf d}=(d_{x}, d_{y})$,
such as those in a Bravais lattice, 
${\cal H}[v, a_{0}(\hat{\bf x}) ]$ and $\Xi$ (with $\ell\, \tilde{\theta} \rightarrow {\bf d}$) 
commute, and 
$\psi$ and $\Xi\, \psi$ belong to the same eigenvalue.
For such periodic systems
magnetic translations\cite{MT_B,MT_Zak} 
play an essential role in classifying
the degeneracy of the eigenmodes, known as the magnetic Bloch bands.

We end this section by referring to 
the standard picture$^{1,14-16,18}$ 
of the integer QHE.
In the presence of disorder each Landau level is turned into a broadened subband. 
The majority of electrons gets localized and electron modes remain extended 
only about the center of the subband spectrum and/or near the sample edges. 
Localized modes cease to carry current 
while a filled subband recovers the same amount of current 
as in the impurity-free case as long as each subband remains distinct.  
The quantized Hall conductance thereby is realized 
when the Femi energy lies in the mobility gap.

\section{Graphene}

The electrons in graphene are described by two-component spinors 
on two inequivalent lattice sites $(A,B)$. 
They acquire a linear spectrum (with velocity $ v_{\rm F} \sim 10^{6}$m/s) 
near the two inequivalent Fermi points $(K,K')$ in momentum space, 
and are described by an effective Hamiltonian of the form,\cite{Semenoff}  
\begin{eqnarray} 
H &=&\int dx dy \, \{ \Psi^{\dag}_{+} {\cal H}_{+}\Psi_{+} + \Psi_{-}^{\dag} {\cal H}_{-}\Psi_{-} \},  \nonumber\\
{\cal H}_{\pm} &=& 
v_{\rm F}\, (\Pi_{1}\sigma^{1}+ \Pi_{2}\sigma^{2} \pm \delta m\, \sigma^{3} ) - eA_{0},
\label{H_GR}
\end{eqnarray}
where $\Pi_{i}= p_{i}+eA_{i}$  [with $i=(1,2)$ or $(x,y)$] 
involve coupling to potentials 
$(A_{i}, A_{0})$ and $\sigma^{i}$ denote Pauli matrices.
The Hamiltonians ${\cal H}_{\pm}$ describe electrons 
at two different valleys $a \in (K,K')$ per spin, and 
$\delta m$ stands for a possible sublattice asymmetry; 
we take $\delta m > 0$, without loss of generality.
Actually, valley asymmetry of a few percent is inferred from experiments\cite{HYYY,CSYL} 
using high-mobility graphene/hexagonal boron nitride (hBN) devices.

Let us place graphene in a uniform magnetic field $B_{z}=B>0$ 
and, as in Sec.~2, 
include also spatially-uniform potentials 
$v(t) = e \ell\, (v_{y} + iv_{x})/\sqrt{2}$ and $v_{0}(t)$.
In the $|n,y_{0}\rangle$ representation, 
the Hamiltonian ${\cal H}_{+}$ at valley $K$ is written as
\begin{equation}
{\cal H}_{+}[v, v_{0}] = \omega_{c} \left(
\begin{array}{ll}
\mu & -Z \!-i v \\
-Z^{\dag} \!+iv^{\dag}& -\mu \\
\end{array}\!
\right) - e v_{0}, 
\label{Hv_gr}
\end{equation}
where $Z= (Y+ iP)/\sqrt{2}$.
Here we have set, along with $\ell \equiv 1/\sqrt{eB}$, 
\begin{equation}
\omega_{c} \equiv \sqrt{2}\, v_{\rm F}/\ell \ \ {\rm and}\ \ 
\mu \equiv  \ell\, \delta m/\sqrt{2}.
\end{equation}

For $v_{\mu}=0$, one can readily diagonalize ${\cal H}_{\pm}$. 
The electron spectrum forms an infinite tower of 
Landau levels of energy 
\begin{equation} 
\epsilon_{n} =s_{n}\,  \omega_{c} \sqrt{|n|+\mu^{2}}
\end{equation}
at each valley (with $s_{n}\equiv  {\rm sgn}[n] = \pm1$), 
labeled by integers $n \in (0,\pm 1, \pm2, \dots)$ and
$y_{0}=\ell^2 p_{x}$, of which only the $n=0$ (zero-mode) levels split in the valley
(hence to be denoted as $n=0_{\mp}$), 
\begin{equation}
\epsilon_{0_{\mp}}= \mp  v_{\rm F}\, \delta m = \mp \omega_{c}\,  \mu   \ \ {\rm for}\ K/K'.
\end{equation}
Thus, for each integer $|n| \equiv N \in (0,1,2, \cdots)$ 
(we use capital letters for the absolute values),
there are in general two modes with $n=\pm N$ (of positive/negative energy) 
at each valley per spin, apart from the $n=0_{\pm}$ modes.

The eigenmodes at valley $K$ are written as 
\begin{equation}
\phi_{n, y_{0}}|^{K} 
= \big( | N\! -\!1, y_{0} \rangle\, b^{n}, | N, y_{0} \rangle\, c^{n} \big)^{\rm t},
\label{psi_n}
\end{equation}
with $(b^{n}, c^{n})^{\rm t}$ given 
by the (normalized) eigenvectors of the reduced (numerical) matrix 
${\cal H}_{+}|_{N}^{\rm red}$
obtained from ${\cal H}_{+}[0,0]$ by replacing $Z, Z^{\dag}\rightarrow \sqrt{N}$.
In explicit form,
\begin{equation} 
(b^{n}, c^{n})= \textstyle{1\over{\sqrt{2}}}\, ( \sqrt{1+ \mu/e_{n} }, -s_{n} \sqrt{1- \mu/e_{n} }),\ \ 
(b^{0_{-}}, c^{0_{-}}) = (0, 1),    
\label{bc_muzero}
\end{equation}
where 
$e_{n}\equiv \epsilon_{n}/\omega_{c}= s_{n}\sqrt{N + \mu^2}$.

 One can pass to another valley $K'$ by simply setting $\mu\rightarrow - \mu$.
Alternatively, note  the relation $\sigma^{3}\, {\cal H}_{-}\sigma^{3} = -{\cal H}_{+}$
which relates the two valleys,
\begin{eqnarray}
&&\phi_{n, y_{0}}|^{K'} = \sigma^3\,  \phi_{-n, y_{0}}|^{K},\ 
\nonumber\\
&&
(\epsilon_{n}, b^{n}, c^{n})|^{K'} = (-\epsilon_{-n}, b^{-n}, -c^{-n})|^{K}.
\label{eh-Conjugation}
\end{eqnarray}
This represents the invariance of $H$ under electron-hole
($e$-$h$) conjugation, i.e., forming another valley 
by interchanging electrons and holes 
$(n \rightarrow -n)$ in a valley. One can also define $e$-$h$ conjugation within a valley
by replacing $\mu\rightarrow -\mu$,
\begin{equation}
(\epsilon_{n}, b^{n}, c^{n}) = (-\epsilon_{-n}, b^{-n}, -c^{-n})|^{\mu\rightarrow -\mu}, 
\end{equation}
in obvious notation, with $n=0 \rightarrow 0_{\mp}$ in valley $K/K'$.

Let us now turn on $(v, v_{0})$.
We expand $\Psi_{\pm}$  in terms of the eigenmodes of ${\cal H}_{\pm}[0,0]$,
\begin{equation}
\Psi_{\kappa} =\int dy_{0} \sum_{n} 
 \left(\!
\begin{array}{l}
| N\! -\!1, y_{0} \rangle\, b^{n}_{\kappa} \\
| N, y_{0} \rangle\, c^{n}_{\kappa} \\
\end{array} \!\!
\right)\,
\psi^{n}_{\kappa}(y_{0}),
\label{Psi_k}
\end{equation}
where $\kappa \in (K,K')$ or $(+,-)$ refers to the valley.  
The Hamiltonian $H$ is then written as
\begin{eqnarray}
H &=&\int\! d y_{0} \sum_{\kappa} 
\psi_{\kappa}^{m \dag}(y_{0}) (\hat{\cal H}_{\kappa}[v,v_{0}] )^{mn} \psi_{\kappa}^{n}(y_{0}),
\nonumber\\
\hat{\cal H}_{+}[v,v_{0}] &=& \omega_{c} \big\{ 
-b\, (Z+ iv)\, c - c\, (Z^{\dag} - iv^{\dag})\, b
\nonumber\\
&&+ \mu\,  (b\, b - c\, c) \big\} -  e\, v_{0}\,  (b\, b+ c\, c), \ \ \ 
\label{hatH_gr}
\end{eqnarray}
where orbital labels $(m,n)$ now run over 
all integers $(0, \pm1,\pm2, \cdots)$. 
[For notational clarity,  we henceforth suppress
obvious valley (and spin) labels, and mainly display $K$-valley expressions.]
Here we have introduced condensed notation: 
For $ \hat{\cal H}[v,v_{0}]^{mn}$ we interpret, e.g., 
\begin{eqnarray}
b\, Z\, c\,  &\rightarrow&\,   b^{m}\, Z^{M-1,N}\, c^{n}, \ 
b\, iv\, c \, \rightarrow\,    iv\,  b^{m}\, 1^{M-1,N} c^{n},\nonumber\\
b\, b\,   &\rightarrow&\,  b^{m}\, 1^{M-1,N-1}\, b^{n}, \ 
c\, c\,  \rightarrow\,   c^{m}\, 1^{M,N}\, c^{n},
\label{condensed_n}
\end{eqnarray}
with $M=|m|$,  $N=|n|$,
 $Z^{M-1,N} \equiv \langle M-1|Z|N\rangle=\sqrt{N}\delta^{M,N}$,
 $1^{M-1,N}\equiv \delta^{M-1,N}$, etc.
Such rules follow from the spinor structure of $\Psi$.
Note that the combination $(b\, b + c\, c)$ in Eq.~(\ref{hatH_gr})
is actually equal to 1 since 
$(bb+cc)^{mn} = (b^{m}b^{n}+c^{m}c^{n} )\, \delta^{MN}=\delta^{mn}$.
For $v\not=0$, $\hat{\cal H}[v,v_{0}]$ is no longer diagonal and
is extended over all sectors of $N= 0,1,2, \cdots$.

Similarly, the charge density 
$\rho_{\bf -p} =
\int d^2 {\bf x}  \sum_{\kappa}\Psi_{\kappa}^{\dag} e^{i{\bf p} \cdot {\bf x}}\, \Psi_{\kappa}$
is rewritten as
\begin{eqnarray}
\rho_{\bf -p} &=& G^{mn}_{\bf p} \int dy_{0}\, \psi^{m \dag} e^{i{\bf p \cdot r } }\, 
\psi^{n} \equiv  G^{mn}_{\bf p}\, R^{mn}_{\bf -p},
\nonumber\\
G^{mn}_{\bf p}
&=& b^{m}\, U^{M-1,N-1}_{\bf p} \, b^{n}
+ c^{m}\, U^{M,N}_{\bf p} \, c^{n}, 
\end{eqnarray}
where 
$U_{\bf p} = e^{ i(p Z^{\dag} + p^{\dag}Z)}$ and
$U^{M,N}_{\bf p} \equiv \langle M| U_{\bf p} | N\rangle$;
$G^{mn}_{\bf p}$ and $R^{mn}_{\bf -p}$ refer to each valley $\kappa$ 
through $b_{\kappa},c_{\kappa}, \psi_{\kappa}$, etc.
Setting $G^{mn}_{\bf p}= \gamma_{\bf p}\, g^{mn}_{\bf p}$, 
one can express  $g^{mn}_{\bf p}$ in terms of polynomials 
$f^{mn}_{\bf p}$ defined in Eq.~(\ref{f_mn}),
\begin{equation}
g^{mn}_{\bf p}
= b^{m}\, f^{M-1,N-1}_{\bf p} \, b^{n}
+ c^{m}\, f^{M,N}_{\bf p} \, c^{n}.
\end{equation}
$e$-$h$ conjugation in Eq.~(\ref{eh-Conjugation}) relates $g^{mn}_{\bf p}$ at the two valleys,
\begin{equation}
g^{mn}_{\bf p}|^{K'} = g^{-m, -n}_{\bf p}|^{K}= g^{m, n}_{\bf p}|^{K; \mu\rightarrow -\mu}.
\end{equation}

Let us now recall that, for each $N=|n|$, the reduced matrix ${\cal H}_{+}|^{\rm red}_{N}$ 
is a real symmetric matrix. Put the associated eigenvectors
${\bf v}_{N} =(b^{N}, c^{N})^{\rm t}$ and ${\bf v}_{-N} =(b^{-N}, c^{-N})^{\rm t}$
into the orthogonal matrix $T=({\bf v}_{N}, {\bf v}_{-N})$.
Obviously the row vectors also form an orthonormal set, which we denote 
as $b \sim (b^{N},b^{-N})^{\rm t}$ and $c \sim (c^{N},c^{-N})^{\rm t}$.  
We write their inner products (e.g., $b\cdot b \equiv b^{N}b^{N}+ b^{-N}b^{-N}$) as
\begin{equation}
b  \cdot b  = c \cdot c =1, \ \ b \cdot c = c \cdot b = 0
\end{equation}
for each $N$ and subsequently extend them to all integers $N$.
In this way, the orbital space $\{ n\}$ is decomposed into 
two subspaces referring to $(b,c)$.
[For the $N=0$ sector one only has $c^{0_{\mp}} =\pm 1$ (and $b\, b=0$);
in most cases $b^{0}=0$ is automatically eliminated via the associated matrix elements
like $c^{m}1^{M,N-1}\, b^{n}$.]
Note that $b\, b$ and $c\, c$, defined in Eq.~(\ref{condensed_n}), 
act as the projection operators,
\begin{eqnarray}
b\,b \cdot b\, b =b\, b,\  c\, c \cdot c\, c = c\, c,\  b\,b+ c\,c = 1.
\end{eqnarray}
One can, of course, verify these properties using the explicit form of $(b,c)$ in Eq.~(\ref{bc_muzero}).  
It will be clear from the above discussion that they are 
a general property of multi-component systems.

Inner products play a role in multiplication, such as
$(b\, U_{\bf p}\, b)^{mj}  (b\, U_{\bf q}\, b)^{jn} 
= b^{m} (U_{\bf p}\, U_{\bf q})^{M-1,N-1}\, b^{n}
= b\, U_{\bf p}\, U_{\bf q}\, b$ and $({\cal O}\, b)\cdot (c\, {\cal O'})=0$.
(For conciseness, we suppress $\lq\lq \cdot$" for an inner product, unless a confusion arises.)
It is now clear that $G_{\bf p}= b\, U_{\bf p}\, b + c\, U_{\bf p}c$ 
enjoys the same composition law as $U_{\bf p}$ in Eq.~(\ref{U_complaw}),
\begin{equation}
G_{\bf p}\,  G_{\bf q} 
= G_{\bf p+q}\,  e^{i {1\over{2}} \ell^2 {\bf p} \times {\bf q}},
{\rm etc.}
\end{equation}
One can even write $G_{\theta}$ in the exponential form 
\begin{eqnarray}
 G_{\theta} &=& b\, U_{\theta}\, b + c\, U_{\theta}\, c 
 =  e^{i(\theta\, {\cal Z}^{\dag}  + \theta^{\dag}{\cal Z})},
 \nonumber\\
 {\cal Z} &\equiv& b\, Z\, b + c\, Z\, c,\ \  
{\cal Z}^{\dag} \equiv b\, Z^{\dag}\, b + c\, Z^{\dag} c,
 \label{Omega_theta}
\end{eqnarray}
with 
$[{\cal Z}, {\cal Z}^{\dag}]=1$; ${\cal Z}^{mn} \propto \delta^{M, N-1}$ and
$({\cal Z}^{\dag})^{mn} \propto \delta^{M, N+1}$ 
thus replace $(Z, Z^{\dag})$ in $U_{\theta}$.

It is now evident that, as in Sec.~2, the rotations 
\begin{equation}
\psi (y_{0}) \rightarrow \psi_{\theta}^{m}(y_{0}) = G^{mn}_{\theta}\, \psi^{n}(y_{0})
\end{equation}
in the orbital space (with $\theta$ common to both valleys)
leave the Coulomb interaction invariant, $V[\rho] = V[\rho_{\theta}]$.
As verified readily, $G_{\theta}$, acting on $\hat{\cal H}$, shifts $v$, 
\begin{equation}
G_{\theta} \hat{\cal H}[v, v_{0}] G_{\theta}^{-1} =\hat{\cal H}[v-\theta, v_{0}].
\label{GHGinv}
\end{equation}

Let us here introduce static fields ${\cal E}_{i}$ by setting
$v_{0} \rightarrow a_{0}({\bf  x}) = v_{0} - x_{j}{\cal E}_{j}$ 
for the reasons that become clear soon.
We denote 
$\hat{\cal H}[v,a_{0}(\hat{\bf x})] \equiv \hat{\cal H}[v,a_{0}({\bf r}); {\cal E}]$, or
\begin{eqnarray}
\hat{\cal H}[v,a_{0}({\bf r}); {\cal E}] 
&=& \epsilon + {\cal  H}_{J} - e a_{0}({\bf  r})  
+e \ell ({\cal E} {\cal Z}^{\dag} + {\cal E}^{\dag}{\cal Z}), \ \ 
\label{HaE} \\
({\cal  H}_{J})^{mn} &=&  i\omega_{c}\,  
(  - b\, v\,  c + c\, v^{\dag}\, b )^{mn},
\label{H_J}
\end{eqnarray}
with $\epsilon \rightarrow \epsilon_{n}\, \delta^{mn}$;
${\cal E} = ({\cal E}_{y}+ i {\cal E}_{x})/\sqrt{2}$.
The full Lagrangian 
\begin{equation}
L=\int\! dy_{0}\, 
\psi^{\dag} \left\{ i\partial_{t}-  \hat{\cal H}[v, a_{0}({\bf r}); {\cal E}]  \right\} \psi
- V[\rho]
\label{L_GR}
\end{equation}
then becomes invariant under the transformation 
$\psi \rightarrow \psi_{\theta} = G_{\theta}\, \psi$ 
and
$(v, a_{0}({\bf r}), {\cal E}) \rightarrow (v^{\theta}, a_{0}^{\theta}({\bf r}), {\cal E}^{\theta})$,
with
\begin{eqnarray}
v^{\theta} &=& v- \theta, \ \ {\cal E}^{\theta} =  {\cal E} - \dot{\theta}/(e\ell),
\nonumber\\
e\, a_{0}^{\theta}({\bf r}) 
&=& e\, a_{0}({\bf r}- \ell \tilde{\theta})  
- i {\textstyle{1\over{2}}} (\theta\, \dot{\theta}^{\dag} - \theta^{\dag} \, \dot{\theta}),
\label{LW_GS_gr}
\end{eqnarray}
where  
$a_{0}({\bf r}- \ell \tilde{\theta}) 
= a_{0}({\bf r}) -  i\ell\, (\theta^{\dag} {\cal E} - \theta\, {\cal E}^{\dag} )$ 
and $\tilde{\theta} \equiv  (\theta_{y}, -\theta_{x})$.
The (gauge-invariant) electric field 
$-\dot{v}/e\ell + {\cal E} = E=(E_{y} + iE_{x})/\sqrt{2}$ thereby
remains invariant.

The present spinor system realizes a long-wavelength gauge symmetry 
with $(v,a_{0}({\bf r}), {\cal E})$.
Setting $\theta=v$ eliminates $v^{\theta}$ but ${\cal E}^{\theta=v} =E$ remains.
It is not possible, unlike in Sec.~2,  to diagonalize the one-body Hamiltonian 
by use of this gauge symmetry alone.
Actually, 
with $\psi_{v}\equiv \psi_{\theta =v}$
and $\hat{\cal H}_{v} \equiv  \hat{\cal H}[ 0 ,a_{0}^{v}({\bf r}); E]$,
one encounters essentially the same structure as ${\cal H}_v$ in Eq.~(\ref{H_v}),
\begin{eqnarray}
L&=&\int\! dy_{0}\, \psi^{\dag}_{v}\, 
\{ i\partial_{t}-  \hat{\cal H}_{v} \}\, \psi_{v}
- V[\rho_{v}],
\nonumber\\
\hat{\cal H}_{v} 
&=&\epsilon  + e\ell (E\, {\cal Z}^{\dag} + E^{\dag}\, {\cal Z} ) - e\, a_{0}^{v}({\bf r}),
\nonumber\\
e\, a_{0}^{v}({\bf r})  
&=& e\, a_{0}({\bf r}- e\ell^2 \tilde{\bf v})  
- i {\textstyle{1\over{2}}} (v\, \dot{v}^{\dag} -v^{\dag} \, \dot{v}), 
\label{psi_v_system}
\end{eqnarray}
where $\tilde{\bf v} \equiv  (v_{y}, -v_{x})$.
From this one can read, as in Sec.~2, the exact Hall conductance 
$\sigma_{xy} = -e^2/h$ per filled level. 
The optical conductance $\sigma_{xy}(\omega)$
is significantly affected by the Coulomb interaction, 
as we will see in Sec.~5.

Each filled level contributes one unit of $-e^2/h$ to $\sigma_{xy}$.
The $\nu = 0$ vacuum state or the infinitely-deep Dirac sea in graphene thus appears 
to carry infinitely large $\sigma_{xy}$, which is unnatural.
The remedy is to handle the Dirac sea carefully, 
assuming a finite depth $n \ge -N_{\rm D}$.

Let us recall that the Chern-Simons term in Eq.~(\ref{psi_v_system}) 
derives from 
$U_{\theta}i\partial U_{-\theta} \ni 
- i{1\over{2}}(\theta\, \dot{\theta}^{\dag} -  \theta^{\dag} \dot{\theta} )\, 
[ {\cal Z},{\cal Z}^{\dag}]$.
In general, $[ {\cal Z},{\cal Z}^{\dag}]^{nn}=1$ 
via four channels of transitions $n \rightarrow k \rightarrow n$
with $|k| =|n| \pm 1$.
For the bottom level $n=-N_{\rm D}$, however, 
one has to omit the $-N_{\rm D}\rightarrow -(N_{\rm D}+1) \rightarrow -N_{\rm D}$ channel 
so that there is no loss of charge.  This yields 
\begin{equation}
[ {\cal Z},{\cal Z}^{\dag}]^{-N_{\rm D},-N_{\rm D}}
= - (N_{\rm D}- 1/2) 
+ O(\mu/\sqrt{N_{\rm D}}).
\end{equation}
The bottom level $n=-N_{\rm D}$ thus  carries 
the amount of current $-(N_{\rm D}  -{1\over{2}})$ times larger.
In consequence, for the $\nu = 0$ state, valley $K$ 
(with $n=0_{-}, -1, \cdots, -N_{\rm D}$ filled) carries $\sigma_{xy}$
equal to $\{N_{\rm D}- (N_{\rm D}- {1\over{2}})\}\,(-e^2/h) = -{1\over{2}}\, e^2/h$ per spin,
while another valley $K'$ (with $n=0_{+}$ empty) 
carries $\sigma_{xy}$ equal to ${1\over{2}}\, e^2/h$ per spin.
This is a manifestation of fermion number fractionalization,
or induced vacuum charge,$^{25-27}$ 
that is traced back to the presence of chiral anomaly in 1+1 dimensions. 
Unfortunately this half unit of conductance is not directly observable 
since a single valley cannot be isolated in equilibrium.  
Thus $\sigma _{xy} = - \nu\,e^2/h$ for a many-body state at integer filling factor $\nu$,
with density $\langle \rho \rangle = \nu\, \bar{\rho}$.

 \section{Conservation Laws}

An alternative yet powerful way to study the response of Hall electrons
is provided by the conservation law associated with the gauge symmetry in Eq.~(\ref{LW_GS_gr}). 
Let us examine how the Lagrangian~(\ref{L_GR}) responds 
to a small $G_{\theta}$ rotation of the electron field, 
$\psi \rightarrow \psi + \delta \psi$ with 
$ \delta \psi = i \{\delta \theta (t) {\cal Z}^{\dag} + \delta \theta^{\dag} (t) {\cal Z}\} \psi$.
The result is the conservation law of the cyclotron-coordinate translation charges 
$\int\! dy_{0}\,  \psi^{\dag} {\cal Z}\psi$ 
or $\int\! dy_{0}\,  \psi^{\dag} {\cal Z}^{\dag} \psi$,
\begin{equation}
-e \ell\, \partial_{t}\! \int\! dy_{0}\,  \psi^{\dag} {\cal Z}\psi 
= \int\! dy_{0}\,  j + i e^2\ell^2{\cal E} \! \int\! dy_{0}\,  \psi^{\dag}\, \psi, {\rm etc.},
\label{X_translation}
\end{equation}
where 
$j \equiv (j_{y} + i j_{x})/\sqrt{2} 
=-e \ell\, \psi^{\dag} (\partial {\cal H}_{J}/\partial v^{\dag})\, \psi
=-i e \ell\, \omega_{c} \psi^{\dag} c\, b\, \psi$ 
stands for the current 
and  
$\int dy_{0} j_{i} = -e\, v_{\rm F} \int d^{2}{\bf x}\,\Psi^{\dag} \sigma^{i} \Psi$ 
in terms of  the field $\Psi$ in the ${\bf x}$ space.
One can also verify this operator equation 
by direct use of the field equation for $\psi$. 
Here we are handling only spatially-averaged quantities 
and this is the reason why  the conservation law takes a simple form 
with no reference to the Coulomb interaction (or, more generally, 
translation-invariant interactions). 
Note that it takes the same form 
as the classical Lorentz equation
\begin{equation}
d{\bf p}/dt = -e ({\bf v}\times {\bf B} + {\bf E}),
\end{equation}
although the correspondence ${\bf p} \propto  \int\! dy_{0}\,  \psi^{\dag} {\cal Z}\psi$
is only suggestive. 
It  is now a simple task to conclude, 
by taking the ground-state expectation value of Eq.~(\ref{X_translation})
for a static configuration with constant ${\cal E}$, 
that 
$\langle j_{x}\rangle = -e^2\ell^2 {\cal E}_{y} \langle \psi^{\dag}\psi\rangle$, 
i.e., 
$\sigma_{xy} = -e^2/h$ per filled level, independent of the Coulomb interaction.

Similarly, the variation  
$\delta \psi = i \{\delta \theta\, r^{\dag}/\ell + \delta \theta^{\dag} r/\ell\}\, \psi$,
associated with magnetic translations $\Xi\, \psi$ with
$\Xi = \exp(i\theta r^{\dag} + \theta^{\dag} r)$ and  
 $r= (r_{y} + ir_{x})/\sqrt{2} = (y_{0} - \ell^2 \partial_{y_{0}})/\sqrt{2}$, 
leads to the conservation law of the center-coordinate translation charge, 
\begin{equation}
-\partial_{t}\! \int\! dy_{0}\,  \psi^{\dag} r\, \psi 
= i {\ell^2\over{\sqrt{2}}} \int\! dy_{0}\, \partial_{y_{0}} (\psi^{\dag} \hat{\cal H}\psi) 
- i e \ell^2{\cal E}\! \int\! dy_{0}\,  \psi^{\dag} \psi ,
\label{r_translation}
\end{equation}
which again is independent of $V[\rho]$.
Note that a total derivative arises from rewriting, e.g.,  
$\delta \psi^{\dag} \propto -i \delta \theta^{\dag}\, (r^{\dag} \psi)^{\dag}$
as $-i \delta \theta^{\dag}\, \psi^{\dag}\, r$ + (a total derivative).
This conservation law shows that, for a static setting, 
the energy difference between the two sample edges $y_{0}= (0, L_{y})$ is
given by the potential difference ${\cal E}_{y}L_{y}$.

Combining Eqs.~(\ref{X_translation}) and~(\ref{r_translation}) 
yields the conservation law associated with the gauge transformation 
$\Lambda= \Xi\, G_{\theta} = e^{i \theta_{j}\hat{x}_{j}/\ell}$,
\begin{equation}
-e \partial_{t}\!\! \int\! dy_{0}\,  \psi^{\dag}(\ell {\cal Z}+ r) \psi 
=\int\! dy_{0}\,   j
+ i {e\ell^2\over{\sqrt{2}}} \int\! dy_{0}\, \partial_{y_{0}} (\psi^{\dag} \hat{\cal H}\psi).
\label{g_translation}
\end{equation}
This equation relates, for a static configuration, 
the total current $\int dy_{0} \langle j_{x}\rangle$ to the energy difference 
$[\langle \psi^{\dag}\hat{\cal H}\psi \rangle]^{L_{y}}_{0}$, and 
has the same content as Eq.~(\ref{deltaH_xi}).
Magnetic translations alone do not directly refer to the current, 
as in Eq.~(\ref{r_translation}), 
but do so when combined with the gauge transformation 
$\Lambda$.
In this way, these conservation laws neatly summarize our analysis in Sec.~3.

The effect of a weak impurity potential $V^{\rm imp}({\bf x})$ 
is also accommodated in these conservation laws.
One can simply replace the Hall potential $-e a_{0}({\bf x})$ 
by a general static potential $V({\bf x}) = -ea_{0}({\bf x}) + V^{\rm imp}({\bf x})$.
Then, in Eqs.~(\ref{X_translation}) and (\ref{r_translation}), 
the $e{\cal E} \int dy_{0}\psi^{\dag} \psi$ term is replaced 
by $-\int d^{2}{\bf x}\, \Psi^{\dag}\{\partial\, V({\bf x})\}\Psi$
with $\partial =(\partial_{y}+ i\partial_{x})/\sqrt{2}$, where we have passed to the field 
$\Psi({\bf x})$ in the ${\bf x}$ space. For a filled level the density 
$\langle \Psi^{\dag}({\bf x})\Psi({\bf x})  \rangle$ becomes uniform and equal to 
$\bar{\rho}$ (per level) since, owing to Fermi statistics, 
an electronic state is either filled or empty 
so that a filled level necessarily attains a uniform density.  
The potential difference is thereby replaced 
by the Hall voltage $\propto \int dy\,  \partial_{y} V({\bf x})$ and 
this leads to the quantized conductance  $\sigma_{xy}= -e^2/h$ per filled level 
when the Fermi energy lies in the mobility gap.

For conventional 2D electrons (of Sec.~2) 
the conservation law~(\ref{X_translation}) retains the same form,
with ${\cal Z} \rightarrow Z$ and 
$j \rightarrow j\equiv  ie\ell \omega_{c} \int  \! dy_{0}\,  \psi^{\dag} (Z+ iv)\,  \psi$;
analogously for Eqs.~(\ref{r_translation}) and (\ref{g_translation}).
In this case, further reduction is possible 
if one notes that the translation charge $\int\! dy_{0}\,  \psi^{\dag} Z\psi$ 
and the current $\int\! dy_{0}\, j$ have essentially the same structure.
Indeed, in view of charge conservation $\partial_{t}\int dy_{0}\, \psi^{\dag}\psi =0$,
one can cast the conservation law in the form 
\begin{equation}
(\omega_{c} - i \partial_{t}) \int\! dy_{0}\, j
= -ie^2\ell^2\omega_{c} \int\! dy_{0}\,  \psi^{\dag} E\, \psi,
\label{conserv_law_two}
\end{equation}
which recovers the optical response~(\ref{current_resp}).
Here this exact optical response holds
for all rotated fields $\psi_{\theta} = U_{\theta}\psi$ 
[with $j = ie\ell \omega_{c} \psi^{\dag}_{\theta}\{Z+i (v-\theta)\}\psi_{\theta}$],
while, at the Lagrangian level, it is made manifest only for $\psi_{\theta'}$
in Eq.~(\ref{L_psi_theta}).

\section{Long-Wavelength Electromagnetic Response}

CR in graphene attracts considerable attention, 
both theoretically$^{28-32}$ 
and experimentally,$^{33-36}$
because one observes a variety of resonance channels and many-body corrections.
The $\psi_{v}$ system in Eq.~(\ref{psi_v_system}), reached via a gauge transformation, 
provides a useful base for deriving, efficiently and in a manifestly gauge-invariant way, 
long-wavelength $({\bf p}=0)$ optical response, which is governed by CR.  
In this section, we calculate such a response and see 
how it is corrected by the Coulomb interaction.

In the $\psi_{v}$ system,
the $V_{E} \equiv E\, {\cal Z}^{\dag} + E^{\dag}\, {\cal Z}$ term causes level mixing.
Let us try to remove this $O(E)$ term from $\hat{\cal H}_{v}$ 
by a general rotation $\psi_{v} \rightarrow \psi'= e^{i \Lambda}\, \psi_{v}$ 
in the orbital space, 
with a hermitian matrix $\Lambda^{mn}= \Lambda^{mn}(t)$.
Consider first the one-body Hamiltonian, which we rewrite as
$H'=\int\! dy_{0}\, \psi'^{\dag}\, \hat{\cal H}_{v}^{(\Lambda)}\, \psi'$
with $\hat{\cal H}_{v}^{(\Lambda)} =e^{i\Lambda}(\hat{\cal H}_{v} - i\partial_{t})e^{-i\Lambda}$,
and calculate the total energy $\langle H' \rangle$ 
for the ground state $|{\rm Gr} \rangle$ of $\psi'$.
The result is 
\begin{eqnarray}
\langle H' \rangle &=& \bar{\rho}\! \int\! d^2{\bf x}\,  
{1\over{2}} \sum_{k,n}(\nu_{n}\! -\nu_{k})\, {\cal H}^{kn} 
= \bar{\rho}\! \int\! d^2{\bf x} \sum_{k>n}(\nu_{n}\! -\nu_{k})\, {\cal H}^{kn},
\nonumber\\
{\cal H}^{k n} &=& \Lambda^{nk} (\epsilon_{k} -\epsilon_{n} - i\partial_{t})\, \Lambda^{kn}
+ i (\Lambda^{nk} V_{E}^{kn} - V_{E}^{nk} \Lambda^{kn} ), 
\end{eqnarray}
where $\nu_{n}=0$ or 1 specifies the occupancy of level $n$.
[Here we have suppressed the $e\, a_{0}^{v}({\bf r})$ term 
which is uncoupled to $\Lambda^{jn}$.]
Note that ${\cal H}^{kn}  = -{\cal H}^{nk} $, up to a total derivative 
$\propto \partial_{t}$;
such (physically inessential) total derivatives will be suppressed from now on.
One can regard $-{\cal H}^{k n}$ as an effective Lagrangian for a field $\Lambda^{kn}$
(of CR in the $n\rightarrow k$ channel) coupled to the Hall field $V_{E}$.

Minimizing $\langle H'\rangle$ with respect to $\Lambda^{kn}$ 
then yields the $O(\alpha^{0})$ optical response 
$\phi_{n\rightarrow k}\propto E^{\dag}E$ 
via the $n \rightarrow k \rightarrow n$ CR transition,
\begin{eqnarray}
\langle H'_{E}\rangle &=&
\bar{\rho}\! \int\! d^2{\bf x} \sum_{k>n} (\nu_{n}-\nu_{k})\, \phi_{n\rightarrow k}. 
\nonumber\\
\phi_{n \rightarrow k}&\equiv&
- V_{E}^{nk} {1\over{\epsilon_{k} - \epsilon_{n} -\omega}}\,  V_{E}^{kn}.
\end{eqnarray}
where $\omega \rightarrow i\partial_{t}$.
For a given ground state one has to sum $\phi_{n\rightarrow k}$ 
over all active resonance channels (with $\nu_{n}-\nu_{k} \not=0$).
The same result is of course reached by the standard perturbation theory.
The present variational method (the sigle-mode approximation)
has the advantage of simplifying and systematizing the higher-order calculations.

Via the rotation $\psi'= e^{i \Lambda}\, \psi_{v}$, the Coulomb interaction 
$V[\rho_{v}] \equiv V[\rho'; \Lambda]$ acquires interaction of 
$O(\alpha\, \Lambda)$, $O(\alpha\, \Lambda^2)$, etc.
The $O(\alpha)$ corrections to ${\cal H}^{kn}$ are extracted from 
the expectation value $\langle V[\rho'; \Lambda]\rangle$.
Actually, one can simply retain 
the diagonal combinations 
$\propto \Delta \epsilon^{k,n} \Lambda^{nk} \Lambda^{kn}$, 
since off-diagonal ones, responsible for mixing among different channels,\footnote{
At zero momentum transfer, there is no mixing between the $\{-(n-1) \rightarrow n\}$ and
$\{-m \rightarrow m-1\}$ transitions induced by an absorption of photons of different circular polarization $E$ and $E^{\dag}$.
} 
eventually contribute to the resonance spectra and 
associated response of $O(\alpha^2)$ or higher. 
It turns out that the $O(\alpha)$ corrections simply modify the CR energy in each channel,  
$\epsilon_{k} -\epsilon_{n} \rightarrow \epsilon^{k, n}_{\rm exc}$,
with
\begin{eqnarray}
\epsilon^{k, n}_{\rm exc} &\equiv& 
\epsilon_{k} -\epsilon_{n}+ \Delta \epsilon^{k,n} = -\epsilon^{n, k}_{\rm exc},
\nonumber \\
\Delta \epsilon^{k,n}&=& \Delta \epsilon_{k} \!- \! \Delta \epsilon_{n} 
- (\nu_{n}\! - \!  \nu_{k})
 \!\sum_{\bf p} v^{\rm C}_{\bf p}\gamma_{\bf p}^{2} g^{kk}_{\bf -p}\, g^{nn}_{\bf p},
\nonumber \\
\Delta \epsilon_{n} &=&  - \sum_{j}  \nu_{j}
\sum_{\bf p} v^{\rm C}_{\bf p}\gamma_{\bf p}^{2} \, |g^{n j}_{\bf p}|^{2}.
\label{deltaEkn}
\end{eqnarray}
The corrections $\Delta \epsilon^{k,n} = -\Delta \epsilon^{n,k}$ consist of 
the exchange self-energies $\Delta \epsilon_{j}$ and 
Coulomb attraction $\propto  v_{\bf p}\gamma_{\bf p}^{2} g^{kk}_{\bf -p}\, g^{nn}_{\bf p}$ 
between the pair of an excited electron and a created hole.
The response $\phi_{n \rightarrow k}$ to $O(\alpha)$ is now cast in the form  
\begin{equation}
\phi_{n \rightarrow k} \equiv  - e^2 \ell^2 
E\,  \Big\{ {{\cal Z}^{nk}({\cal Z}^{\dag} )^{kn}\over{\epsilon^{k, n}_{\rm exc} + \omega}}
+ {({\cal Z}^{\dag} )^{nk} {\cal Z}^{kn}\over{\epsilon^{k, n}_{\rm exc}- \omega}}
 \Big\}\, E^{\dag};
\end{equation}
the numerators are explicitly written as
\begin{equation}
({\cal Z})^{nk} ({\cal Z}^{\dag})^{kn}
=\delta^{K, N+1} {1\over{4}}\, (e_{k} + e_{n})^2 
(1+{\mu\over{e_{k}}})(1-{\mu\over{e_{n}}}). 
\label{ZZdag_GR}
\end{equation}

The long-wavelength (${\bf p}=0$) response in general takes the form
\begin{equation}
\langle H'_{E}\rangle =- \int\! d^2{\bf x}\,\Big\{ 
{1\over{2}}\, E_{i}\,  \alpha_{e}(\omega)\,  E_{i}
+ v_{x} \Delta \sigma_{xy}(\omega) \dot{v}_{y} \Big\}.
\end{equation}
Of $\phi_{n\rightarrow k}$, terms even in $\omega$ contribute 
to the electric susceptibility $\alpha_{e}(\omega)$ and those odd in $\omega$ 
to the frequency dependence of the optical Hall conductivity 
$\Delta \sigma_{xy}(\omega)= \sigma_{xy}(\omega) -\sigma_{xy}(0)$.
The response depends on the filling factor $\nu$ of the ground state, and
we thus specify it by referring to the uppermost filled level $n_{\rm f}$ in each valley. 
For clarity, we focus, in what follows, our attention on the following cases 
of integer filling supporting a distinct mobility gap:
(i) When a sizable Landau gap is present, $n_{\rm f}$ is common to both valleys, 
with total filling factor $\nu= 4 n_{\rm f} + 2=(-6, -2, 2, 6, \dots)$ for $n_{\rm f} = (-2, -1, 0,1, \dots)$.
(ii) With appreciable breaking $\mu$, the $\nu=0$ neutral state also develops 
a band gap, acquiring the valley content $(n_{\rm f}|^{K},\,  n_{\rm f}|^{K'}) = (0,-1)$.

There arises a variety of CR channels in graphene.
Unlike in conventional 2D systems, 
the filled valence band always supports
infinitely many active interband channels, 
such as $\{ -(n-1) \rightarrow n$ and $-n \rightarrow n-1\} \equiv T_{n}$ for $n=1,2,\cdots$,
with the response 
\begin{eqnarray}
 \phi^{+}_{n}(\omega) &\equiv& \phi_{-(n-1)\rightarrow n} 
 =-  {1\over{4}}\, e^2\ell^2 \, 
 E\, {N_{n}(\mu) \over{\epsilon^{+}_{n}+ \omega}} \, E^{\dag}, 
 \nonumber\\
 \phi^{-}_{n}(\omega) &\equiv&  \phi_{-n \rightarrow n-1}
 = -  {1\over{4}}\, e^2\ell^2 \, E\,  
  {N_{n}(-\mu)\over{\epsilon^{-}_{n} - \omega}}\, E^{\dag},
 \nonumber\\
 N_{n}(\mu) &=& (e_{n}  -e_{n-1})^2(1+ \mu/e_{n-1}) (1+ \mu/e_{n}),
\end{eqnarray}
at valley $K$,
where $\epsilon^{+}_{n} \equiv \epsilon_{\rm exc}^{n.-(n-1)}$ and
$\epsilon^{-}_{n} \equiv \epsilon_{\rm exc}^{n-1, -n}$ for short.

Interestingly, these intrerband channels of $T_{n}$ are simultaneously active 
over the interval of $-(n-1) \lesssim n_{\rm f} \lesssim n$ 
or total filling factor $|\nu| \lesssim 4n-2$.
They have the same spectra 
$\epsilon_{n}^{\pm} \rightarrow \epsilon_{n} + \epsilon_{n-1}$ 
for $\alpha = 0$, and are intimately related, for $\alpha \not= 0$, 
between the two valleys (or within a valley) via $e$-$h$ conjugation.
Obviously, via conjugation (i.e., $n \leftrightarrow -n$ and $K\leftrightarrow K'$),
the ground state of filling factor $\nu$ turns into one of filling factor $-\nu$,
and valley $K$ with $n_{\rm f}=m$ turns into valley $K'$ with $n_{\rm f}=-m-1$
(and vice versa).
The CR channels $n \leftarrow -j$ and $j \leftarrow -n$ are thereby interchanged and, 
as shown by examining the explicit form of $\Delta \epsilon^{k,n}$ in Eq.~(\ref{deltaEkn}), 
they share the same spectra at the conjugated valleys,\cite{KS_eh}
 \begin{equation}
\epsilon_{\rm exc}^{n \leftarrow -j}|_{n_{\rm f}=m}^{K} \!
= \epsilon_{\rm exc}^{j \leftarrow -n}|^{K'}_{n_{\rm f}=-m-1}
=\epsilon_{\rm exc}^{j \leftarrow -n}|_{n_{\rm f}=-m-1}^{K;\mu\rightarrow -\mu}.
\end{equation}
For $T_{n}$, in particular, the conjugate valleys have 
essentially the same spectra  $\{\epsilon_{n}^{\pm}\}$,
\begin{equation}
\{\epsilon^{+}_{n}, \epsilon^{-}_{n}\}|^{K}_{\nu} 
= \{\epsilon^{-}_{n}, \epsilon^{+}_{n}\}|^{K'}_{-\nu},
\end{equation}
and, as a result,  the associated responses are mutually related, 
\begin{equation}
\phi_{n}^{\pm} (\omega)|_{\nu}^{K} = \phi_{n}^{\mp} (-\omega)|_{-\nu}^{K'},\ \
\phi_{n}^{\pm} (\omega)|_{\nu}^{K'} =  \phi_{n}^{\mp} (-\omega)|_{-\nu}^{K},
\label{response_KvsKp}
\end{equation}
where $|_{\pm\nu}$ refers to the ground state of filling factor $\pm \nu$.
Similarly, intraband channels $n-1\rightarrow n$ and $-n \rightarrow -(n-1)$ 
also form an $e$-$h$ conjugate pair (though not simultaneously observable) 
and obey the same relations.
Consequently, the ground states of filling $\pm \nu$, in general, support 
essentially the same ($K+K'$) CR spectra
\begin{equation}
\{\epsilon^{+}_{n}, \epsilon^{-}_{n}\}|^{K}_{\nu} 
\oplus \{\epsilon^{+}_{n}, \epsilon^{-}_{n}\}|^{K'}_{\nu} 
= \{\epsilon^{+}_{n}, \epsilon^{-}_{n}\}|^{K}_{\nu}\oplus 
 \{\epsilon^{-}_{n}, \epsilon^{+}_{n}\}|^{K}_{-\nu},
 \label{spec_KvsKp}
\end{equation}
and the same ($K+K'$) response of the form
\begin{equation}
R(\omega)|_{\nu} = R(-\omega)|_{-\nu},  
\label{R_omega_nu}
\end{equation}
as seen, e.g., from the relations
\begin{equation}
\phi_{n}^{\pm} (\omega)|_{\nu}^{K} +\phi_{n}^{\mp} (\omega)|_{\nu}^{K'}
= \phi_{n}^{\pm} (\omega)|_{\nu}^{K} +\phi_{n}^{\pm} (-\omega)|_{-\nu}^{K}.
\end{equation}
Equation~(\ref{R_omega_nu}) thus reveals the general features of $\sigma_{xy}(\omega)$
and $ \alpha_{e}(\omega)$,
\begin{equation}
\sigma_{xy}(\omega)|_{\nu} =- \sigma_{xy}(\omega)|_{-\nu},\ \ 
\alpha_{e}(\omega)|_{\nu} = \alpha_{e}(\omega)|_{-\nu}.
\label{e-h-conj-response}
\end{equation}
The optical conductivity naturally vanishes, $\sigma_{xy}(\omega)=0$, 
for the $\nu=0$ state which is $e$-$h$ self-conjugate.
This vacuum state, on the other hand, 
acquires, as a response of the filled valence band, 
the electric susceptibility,
\begin{eqnarray}
\alpha^{\rm vac}_{e}(\omega)
&=&{1\over{2}}g_{\rm s}\,\bar{\rho} \, e^2\ell^2
\sum_{n=1}^{\infty}\big\{D^{+}(\omega) + D^{-}(\omega)\big\},
\\
D^{\pm}(\omega) &=& N_{n}(\pm \mu)\, \epsilon^{\pm}_{n}/
\{ (\epsilon^{\pm}_{n})^2- \omega^2\},
\end{eqnarray}
where $g_{\rm s}=2$ counts the spin degrees of freedom. 
In the $(\omega, \alpha,\mu) \rightarrow 0$ limit, 
\begin{equation}
\alpha_{e}^{\rm vac}(0)
\stackrel{ \alpha,\mu \rightarrow 0}{=}
{g_{\rm s}\,e^2\over{2\pi\, \omega_{c}}}\,  
\sum_{n=1}^{\infty} {1\over{(\sqrt{n+1} + \sqrt{n})^3}}
\approx {g_{\rm s}\, e^2\over{2\pi\, \omega_{c}}}\,(1.247).
\end{equation}
recovers an earlier result.\cite{KS_emresp}

In retrospect, the $\nu$-dependent features of the CR spectra and response 
in Eqs.~(\ref{spec_KvsKp}) and (\ref{e-h-conj-response})
are what one would naturally expect on the basis of $e$-$h$ conjugation.
They have a special consequence for the interband channels 
$T_{n}$, which 
are active over the range $|\nu| \lesssim 4n-2$  ($n\ge 1$).
The excitation spectrum of each $T_{n}$, 
when observed under fixed magnetic field $B$ over such a finite range of $\nu$, 
will show a profile symmetric in $\nu$ about $\nu=0$.
Such features of many-body corrections are indeed seen in a rather recent observation,
by Russell  et al.\cite{RZTW}, of CR spectra for $T_{1}$ - $T_{6}$ 
in high-mobility hBN-encapsulated graphene.
In this way, interband CR in graphene provides a ground for studying the interaction effects.

It is enlightening to examine how the infinitely-deep valence band affects 
the optical conductivity $\sigma_{xy}(\omega)$.
In the absence of interaction $(\alpha \rightarrow 0)$, 
the excitation spectra $\epsilon_{n}^{\pm} \rightarrow e_{n} + e_{n-1}$ 
have no reference to $\nu$, and 
$\phi_{n}^{\pm} (\omega)|^{K} +\phi_{n}^{\mp} (\omega)|^{K'}\rightarrow$ even in $\omega$;
the filled valence band thus does not contibute to $\sigma_{xy}(\omega)$, 
and only the intraband channels do.
When the Coulomb interaction is turned on ($\alpha \not= 0$), 
in contrast,  the filled valence band does contribute 
to $\Delta \sigma_{xy}(\omega)$ for $\nu\not=0$ 
(because one can verify that $\epsilon_{n}^{-}|^{K'}_{\nu}
= \epsilon_{n}^{+}|^{K}_{-\nu} = \epsilon_{n}^{+}|^{K}_{\nu}  + O(\alpha)$ 
for $\nu\not=0$).  
Here we see explicitly that, unlike Hall conductance $\sigma_{xy}(0)$, 
the optical response $\Delta \sigma_{xy}(\omega)$ is sensitive,
through its $(\omega, \nu)$ dependence, to many-body corrections.

\section{Summary and Discussion}

In this paper 
we have studied electromagnetic response of 2D electrons in a magnetic field
and pointed out  that their response, 
via  spatially-uniform potentials $v_{\mu}$ and fields ${\cal E}_{i}$, 
enjoys a long-wavelength gauge symmetry 
associated with cyclotron motion of electrons.
This gauge symmetry leaves the Coulomb interaction invariant and 
naturally explains why some such long-wavelength response 
as the Hall conductance and cyclotron resonance, 
under certain circumstances, appears insensitive to the interaction.

Special attention has been paid to two types of translations in a magnetic field, 
those ($U_{\theta}$ or $G_{\theta}$) in cyclotron (or relative) coordinates ${\bf X}$ 
and those $(\Xi)$ in center coordinates ${\bf r}$. 
They arise as a projection to the orbital space $\{n\}$ and 
to the center space $\{ {\bf r}\}$, respectively, 
of electromagnetic gauge transformations 
$\Lambda = e^{i\theta_{j} \hat{x}_{j}/\ell}= U_{\theta}\, \Xi$.
The former thus serve to diagonalize the response in $\{n\}$ 
while the latter shift the system spatially in $\{{\bf r}\}$, 
and their actions are related via gauge transformations.
The basic relations between long-wavelength response and translations, 
as well as their insensitivity to the Coulomb interaction, 
are best revealed by the conservation laws 
associated with $U_{\theta}$, $\Xi$ and $\Lambda$, as shown in Sec.~5.
Magnetic translations play a key role in clarifying the effect of disorder and localization
in the QHE, i.e., the immobility of localized electron modes, as discussed in Sec.~3.
For practical calculations of response, 
it is advantageous to handle a suitable 
$U_{\theta}$-transformed form of the Hamiltonian, 
as we have seen in several examples.

The presence of a long-wavelength gauge symmetry directly leads to 
a universal value of the Hall conductance $\sigma_{xy} = -e^2/h$ 
per filled Landau level for 2D electrons. 
The way it is realized, however, is different for Dirac electrons in graphene
and conventional 2D electrons.  
The difference comes from the fact that, for the latter, 
the (spatially-averaged) current operator happens to act as the relative-coordinate translation charge.
As a result, the gauge symmetry and associated conservation law 
become more restrictive for the conventional electrons, 
leading to an exact optical conductance $\sigma_{xy}(\omega)$, 
as implied by Kohn's theorem.

With reduced observable degrees of freedom (i.e., long wavelengths here), 
2D electron systems develop a new gauge symmetry, as we have seen.
Such a viewpoint of an emerging symmetry will be a useful lesson 
from the present paper.

\newpage


\section*{References}
\vspace*{3pt}

\begin{thebibliography}{99}

\bibitem{PG} R. E. Prange and S. M. Girvin (eds.), 
{\it The Quantum Hall effect}
(Springer-Verlag, Berlin, 1987).

\bibitem{NG} K.~S.~Novoselov, A.~K. Geim, S.~V.~Morozov, D.~Jiang,
 M.~I.~Katsnelson, I.~V.~Grigorieva, S.~V.~Dubonos, and 
A.~A.~Firsov, {\it Nature (London)} {\bf 438}, 197 (2005).

\bibitem{ZTSK} Y. Zhang, Y.-W. Tan, H. L. Stormer, and P. Kim, 
{\it Nature (London)} {\bf 438}, 201 (2005).

\bibitem{GN_rev} A. K. Geim and K. S. Novoselov, {\it Nat. Mater.} {\bf 6}, 183 (2007).
 
\bibitem{TSG}
 D.~C.~Tsui, H.~L.~Stormer, and A.~C.~Gossard, 
{\it Phys. Rev. Lett.} {\bf 48}, 1559 (1982). 

\bibitem{Laughlin_wf} R. B. Laughlin, 
{\it Phys. Rev. Lett.} {\bf 50}, 1395 (1983). 

\bibitem{KH} 
C. Kallin and B. I. Halperin,  
{\it Phys. Rev. B} {\bf 30}, 5655 (1984).

\bibitem{GMP}  S. M. Girvin, A. H. MacDonald, and P. M. Platzman,
{\it Phys. Rev. B} {\bf 33}, 2481 (1986).

\bibitem{MZ} A. H. MacDonald and S.-C. Zhang,
{\it Phys. Rev. B} {\bf 49}, 17208 (1994).

\bibitem{Moon} 
K. Moon, H. Mori, K. Yang, S.M. Girvin, A.H. MacDonald, 
L. Zheng, D. Yoshioka, and S.-C. Zhang,
{\it Phys. Rev. B} {\bf 51}, 5138 (1995).

\bibitem{AA} K. Asano and T. Ando, {\it Phys. Rev. B} {\bf  58}, 1485 (1998).

\bibitem{RFG} R. Rold\'an, J. N, Fuchs, and M. O. Goerbig, 
{\it Phys. Rev. B} {\bf  82}, 205418 (2010).

\bibitem{Klitz} K. von Klitzing, G. Gorda, and M. Pepper, 
{\it Phys. Rev. Lett.} {\bf 45}, 494 (1980).  

\bibitem{AAndo} H. Aoki and T. Ando, 
{\it Solid State Commun.} {\bf 38}, 1079 (1981). 

\bibitem{Pra} R. E. Prange, {\it Phys. Rev. B} {\bf 23}, 4802 (1981). 

\bibitem{Laughlin_qhe} R. B. Laughlin, {\it Phys. Rev. B} {\bf 23}, 5632 (1981). 

\bibitem{Kohn} W. Kohn, {\it Phys. Rev.} {\bf 123}, 1242 (1961).  

\bibitem{KS_gauge}  K. Shizuya, {\it Phys. Rev. B} {\bf 45},  11 143 (1992); 
{\it ibid. B} {\bf 52},  2747 (1995).

\bibitem{MT_B}  E. Brown, {\it Phys. Rev.} {\bf 133},  A1038 (1964).

\bibitem{MT_Zak}  J. Zak, {\it Phys. Rev.} {\bf 134},  A1602,  (1964); 
{\it ibid.} {\bf 134},  A1607 (1964).

\bibitem{Berry} 
M.~V.~Berry, {\it Proc. R. Soc. London, Ser. A} {\bf 392}, 45 (1984).

\bibitem{Semenoff}  G. W. Semenoff, {\it Phys. Rev. Lett.} {\bf 53}, 2449 (1984).

\bibitem{HYYY} B. Hunt, J. D. Sanchez-Yamagishi, A. F. Young, M. Yankowitz,
B.~J. Leroy, K.~Watanabe, T.~Taniguchi, P. Moon, M. Koshino,
P. Jarillo-Herrero, and R. C. Ashoori, 
{\it Science} {\bf 340}, 1427 (2013).

\bibitem{CSYL} Z.-G. Chen, Z. Shi, W. Yang, X. Lu, Y. Lai, H. Yan, F. Wang,
G. Zhang, and Z. Li, 
{\it Nat. Commun.}~{\bf 5}, 4461 (2014).


\bibitem{J}
R. Jackiw and C. Rebbi, {\it Phys. Rev. D} {\bf 13}, 3398 (1976).

\bibitem{NS} A.~J. Niemi and G.~W.~Semenoff, 
{\it Phys. Rev. Lett.} {\bf 51}, 2077 (1983).

\bibitem{Red} A. N. Redlich, {\it Phys. Rev. Lett.} {\bf 52}, 18 (1984);

\bibitem{AbF} 
D.~S.~L.~Abergel and V.~I.~Fal'ko, {\it Phys. Rev. B} {\bf 75}, 155430 (2007).

\bibitem{IWF} A. Iyengar, J. Wang, H. A. Fertig, and L. Brey, {\it Phys. Rev. B} {\bf 75},
125430 (2007).

\bibitem{BM} Yu. A. Bychkov and G. Martinez, {\it Phys. Rev. B} {\bf 77}, 125417 (2008).

\bibitem{KS_CR}  K. Shizuya, {\it Phys. Rev. B} {\bf 81}, 075407 (2010).

\bibitem{KS_eh}  K. Shizuya, {\it Phys. Rev. B} {\bf 98}, 115419 (2018);
{\it Int. J. Mod. Phys. B} {\bf 31}, 1750176 (2017).

\bibitem{JHTWS} Z. Jiang, E.~A. Henriksen, L.~C.~Tung, Y.-J.~Wang, M.~E.~Schwartz,
M.~Y. Han, P.~Kim, and H.~L. Stormer, 
{\it Phys. Rev. Lett.} {\bf 98}, 197403 (2007).   

\bibitem{DCNN}R. S. Deacon, K.-C. Chuang, R.~J. Nicholas, 
K.~S. Novoselov, and A.~K. Geim, 
{\it Phys. Rev. B} {\bf 76}, 081406(R) (2007).

\bibitem{HCJL} E. A. Henriksen, P. Cadden-Zimansky, Z. Jiang, Z. Q. Li, L.-C. Tung, 
M. E. Schwartz, M. Takita, Y.-J. Wang, P. Kim, and H. L. Stormer, 
{\it Phys. Rev. Lett.} {\bf 104}, 067404 (2010).


\bibitem{RZTW} B. J. Russell, B. Zhou, T. Taniguchi, K. Watanabe, and E. A.
Henriksen, 
{\it Phys. Rev. Lett.} {\bf 120}, 047401 (2018).

\bibitem{KS_emresp}  K. Shizuya, {\it Phys. Rev. B} {\bf 75}, 245417 (2007).


\end{thebibliography}
\end{document}